\documentclass[11pt,twocolumn,twoside]{article}

\usepackage[paperwidth=205mm,paperheight=285mm,
            left=15mm,right=15mm,top=30mm,bottom=28mm,
            headheight=12pt,headsep=5mm,footskip=16pt,
            columnsep=5mm]{geometry}

\usepackage[T1]{fontenc}
\usepackage[utf8]{inputenc}
\usepackage{mathptmx}          
\usepackage{courier}           
\usepackage{amsmath,amssymb}
\usepackage{microtype}
\usepackage{cuted}            

\makeatletter
\renewcommand\normalsize{\@setfontsize\normalsize{11}{11.69}}
\makeatother
\normalsize
\usepackage{indentfirst}
\usepackage{titlesec}
\titleformat{\section}{\normalfont\bfseries\centering}{\thesection.}{0.5em}{\MakeUppercase}
\titlespacing*{\section}{0pt}{9pt plus 2pt minus 2pt}{5pt}
\titleformat{\subsection}{\normalfont\bfseries\centering}{}{0pt}{}
\titlespacing*{\subsection}{0pt}{8pt}{4pt}

\usepackage{fancyhdr}
\fancypagestyle{plain}{\fancyhf{}\fancyfoot[C]{\small\thepage}%
  }

\usepackage{graphicx}

\usepackage[font=footnotesize,labelsep=period,labelfont=bf,justification=justified,
            singlelinecheck=false,skip=5pt]{caption}
\usepackage[hidelinks]{hyperref}

\newcommand{\StI}{\mathrm{I}}
\newcommand{\StQ}{\mathrm{Q}}
\newcommand{\StU}{\mathrm{U}}
\newcommand{\StV}{\mathrm{V}}
\newcommand{\IIV}{I_{\mathrm{I}\pm\mathrm{V}}}
\newcommand{\BIV}{B_{\mathrm{IV}}}
\newcommand{\BIVl}[1]{B_{\mathrm{IV},#1}}
\newcommand{\BIpmV}{B_{\mathrm{I}\pm\mathrm{V}}}
\newcommand{\BME}{B_{\mathrm{ME}}}
\newcommand{\BMEpar}{B_{\mathrm{ME},\parallel}}
\newcommand{\Bpar}{B_{\parallel}}
\newcommand{\kt}{k_{12}}
\newcommand{\kme}{k_{\mathrm{MEIV}}}
\newcommand{\WI}{W_{\mathrm{I}}}
\newcommand{\WP}{W_{P}}
\newcommand{\Ptot}{P_{\mathrm{tot}}}
\newcommand{\Ic}{I_{\mathrm{C}}}
\newcommand{\IQS}{I_{\mathrm{QS}}}
\newcommand{\lIV}{\lambda_{\mathrm{I}\pm\mathrm{V}}}
\newcommand{\FeI}{Fe\,\textsc{i}}
\newcommand{\mA}{m\text{\AA}}
\newcommand{\flux}{Mx\,cm$^{-2}$}
\newcommand{\degr}{^{\circ}}

\newcommand{\Id}[1]{\textbf{\small\texttt{#1}}}

\newenvironment{journalrefs}
  {\begin{list}{}{%
     \setlength{\labelwidth}{1.6em}%
     \setlength{\labelsep}{0.4em}%
     \setlength{\leftmargin}{2em}%
     \setlength{\itemindent}{0pt}%
     \setlength{\itemsep}{3pt plus 1pt}%
     \setlength{\parsep}{0pt}%
     \setlength{\topsep}{4pt}%
     \small
     \renewcommand{\makelabel}[1]{\textemdash\hfill}}}
  {\end{list}}

\begin{document}

\twocolumn[%
  \vspace*{-6mm}
  {\centering
   \bfseries\fontsize{14}{16.5}\selectfont
   \MakeUppercase{Quiet-Sun magnetic field configuration \\ inferred from 
   the ratio of longitudinal fields \\ measured in the \FeI{} $\lambda$6301 
   and $\lambda$6302~\AA{} lines}\par}

  \vspace{5mm}
  {\centering\bfseries\fontsize{12}{14}\selectfont
   S.~G.~Mozharovsky\par}

  \vspace{2.5mm}
  {\centering\itshape\fontsize{10}{12}\selectfont
   Institute of Applied Astronomy RAS, Saint Petersburg, Russia\par}

  \vspace{1.5mm}
  {\centering\itshape\small
   e-mail: mozharovskys@mail.ru\par}
  {\centering\small
   ORCID: 0000-0002-5552-8254\par}

  \vspace{3mm}
  {\centering\small
   Received by the editorial office on February 24, 2025.
   After revision in 2025.\par
   Accepted for publication in 2025.\par}

  \vspace{5mm}
  {\small\leftskip=10mm \rightskip=10mm
   \noindent
   The estimate of the longitudinal magnetic field obtained from the
   Hinode spectropolarimeter data will be different if it is determined
   in different ways. In particular, the values obtained by the Center
   Of Gravity (COG) method for the \FeI{} $\lambda$6301 and 6302~\AA{}
   lines do not coincide. Some of the differences are due to the
   different inaccuracies of the COG method for each of the two lines.
   However, some of the differences can be explained by the fact that
   the response functions of these lines to changes in the magnetic
   field are formed at different heights. This should be accompanied by
   a certain morphological structure of magnetic fields. It consists in
   the fact that most of the magnetic configurations of the quiet
   photosphere are similar to each other. To a rough approximation,
   their structure can be compared with the structure of a sunspot ---
   there is a central region with the strongest field, close to
   vertical, and slopes --- an analog of penumbra, where the field
   weakens and\,/\,or turns to horizontal. This turn occurs at the
   heights of the greatest response of the \FeI{} $\lambda$6301 and
   6302~\AA{} lines to changes in the magnetic field with a turn length
   of about tens of kilometers.\par}

  \vspace{3.5mm}
  {\small\leftskip=10mm \rightskip=10mm
   \noindent\textit{Keywords:} Hinode SOT/SP, longitudinal magnetic
   fields, Quiet Sun, COG method\par}

  \vspace{7mm}

  \vfill
  \raggedright
  {\small DOI: 10.7868/S3034502225070128 \\
  Published in Geomagnetism and Aeronomy (Russian edition), 
  2025, Vol. 65, No. 7, pp. 1073--1082\par}

]
\thispagestyle{plain}


\section{Introduction}

A well-known problem is the discrepancy between magnetic field values
obtained using different measurement methods. We illustrate these
differences graphically (see Fig.~1), using the extensive data collected
by the Hinode spectropolarimeter.

\begin{figure*}[!t]
  \centering
  \includegraphics[width=\textwidth]{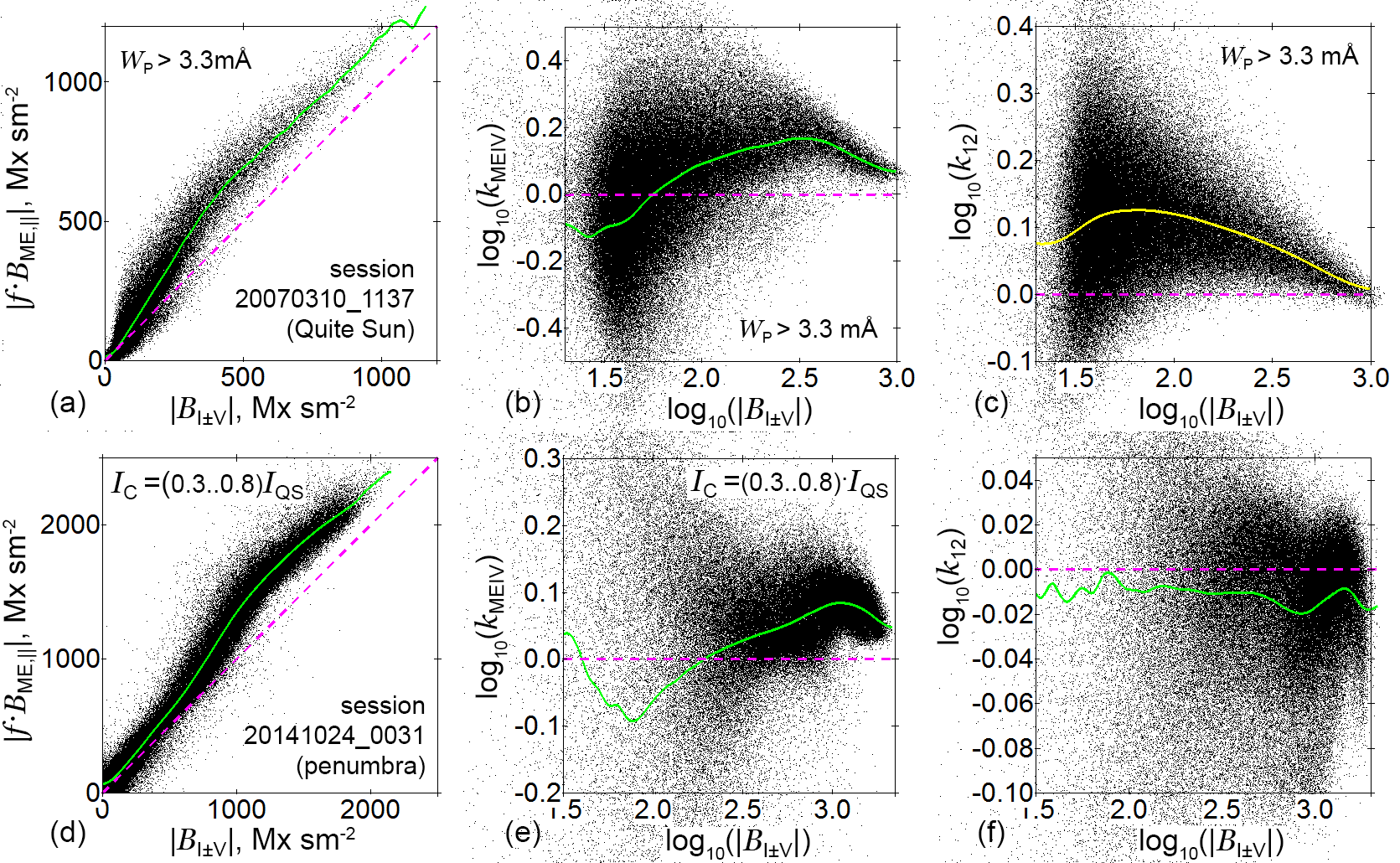}
  \caption{The top row -- data from session 20070310\_1137 (quiet
  photosphere). The bottom row -- session 20141024\_0031 (data taken for
  penumbral regions). The diagrams (a, b, d, e) compare the longitudinal
  magnetic flux densities $f\cdot\Bpar$ obtained using the ME-inversion
  method with the values of $\BIV$ obtained using the COG method; the pair
  of diagrams (a) and (b) as well as the pair (d) and (e) display the same
  data, only the physical axes differ. The diagrams (c, f) show a
  comparison of the values of $\BIVl{6301}$ and $\BIVl{6302}$, obtained
  using the COG method.}
  \label{fig:1}
\end{figure*}

First, let us compare the longitudinal field values in the quiet
photosphere measured by the Milne--Eddington model-based inversion
(ME inversion), which we will denote as $\BME$, with the values obtained
using the center of gravity (COG) method for the Stokes profiles $\StI+\StV$
and $\StI-\StV$, which we will denote as $\BIV$; see Figs.~1(a) and~1(b) for the
quiet photosphere, and 1(d) and~1(e) for the penumbra. 
It should be noted that when we discuss the magnetic field derived from ME 
inversion, we have both the field strength (measured in Gauss) and the 
magnetic flux density, which is obtained by multiplying the field strength 
by the filling factor and is measured in Gauss (or Maxwells per square 
centimeter).
However, when we refer to the field measured by the COG method, it is 
always averaged over the observed region. That is, it is averaged over 
both the spectrograph aperture area and the height range within which the 
spectral lines are formed. For simplicity, we often call this quantity the
field strength, but in essence, it represents the average magnetic flux 
density. 

The differences
between $\BMEpar$ and $\BIV$ are not large; some of them can be explained
by inaccuracies in the COG method (see Section~3). Obviously, the response
functions, in particular the effective response heights for the COG and
ME-inversion methods, are different. However, this difference is due to
such a large number of factors that it is not yet possible to analyze it
and relate it to the physical conditions in the photosphere. Secondly, let
us compare the longitudinal field values measured using the COG method
for the \FeI{} $\lambda$6301 and $\lambda$6302~\AA{} lines separately. For
the penumbra (Fig.~1(f)), the differences are minimal and related to the
variation in the application of the COG method to different lines.
However, in the case of a quiet photosphere (Fig.~1(c)) the difference at
first glance requires a sharp increase in the longitudinal field with
height. The second way to explain the picture in Fig.~1(c) is to use a
model with a filling factor $f$, in which the magnetic field occupies only
a small part of the area. The required ratio of measured values
$\kt=\BIVl{6301}/\BIVl{6302}$ can be explained by the difference between
the Stokes profiles $\StI$ in magnetic and non-magnetic elements mixed in the
spectrograph aperture [Domínguez Cerdeña et al., 2003; Botygina et al.,
2017].

In Fig.~1, the data from two observation sessions are presented in the top
and bottom rows. The top row shows a typical session during which a quiet
Sun is observed. The map of this session has the maximum possible size for
the SP Hinode. In the bottom row, one of the sessions is shown where a
significant portion of the area is occupied by the penumbra of a complex
sunspot group.

If the measurement methods were error-free, their results would be the
same. The differences indicate either possible errors in the methods, or
that we receive data from different volumes of solar plasma. For example,
from volumes lying at different heights. Either in different regions of
the aperture area, when the source of the Stokes $\StV$ profile is located
within the region with a magnetic field, while the source of the $\StI$
profile is spread over the entire area. So, according to Figs.~1(a)
and~1(b), in a quiet photosphere, the flux density ratios
$\kme = f\cdot\BMEpar/\BIV$ vary from 0.7 for $\BIV = 30$~\flux{} up to 1.5
at $\BIV = 350$~\flux{}. In the penumbra (Figs.~1(d,\,e)) $\kme$ reach 0.8
and 1.2 at $\BIV = 60$ and 1100~\flux{}, respectively.

The differences in the longitudinal field, obtained using the COG method
for two different lines, are more interesting. Although the differences
are small in the penumbra (Fig.~1(f)) -- namely, the
$\kt=\BIVl{6301}/\BIVl{6302}$ differs from unity by less than 5\% and can
be attributed to the differing accuracy of the method for these lines --
in the quiet photosphere (Fig.~1(c)) the value of $\kt$ reaches on average
1.3 for $\BIV$ 50--80~\flux{} ($\log_{10}(\kt)\approx 0.12$). The value
$\kt = 1.3$ was obtained according to SP/Hinode data also in studies by
other authors [Botygina et al., 2017]. In the paper [Domínguez Cerdeña et
al., 2003] $\kt = 1.25 \pm 0.14$.

\section{Observation data}

\begin{figure*}[!t]
  \centering
  \includegraphics[width=\textwidth]{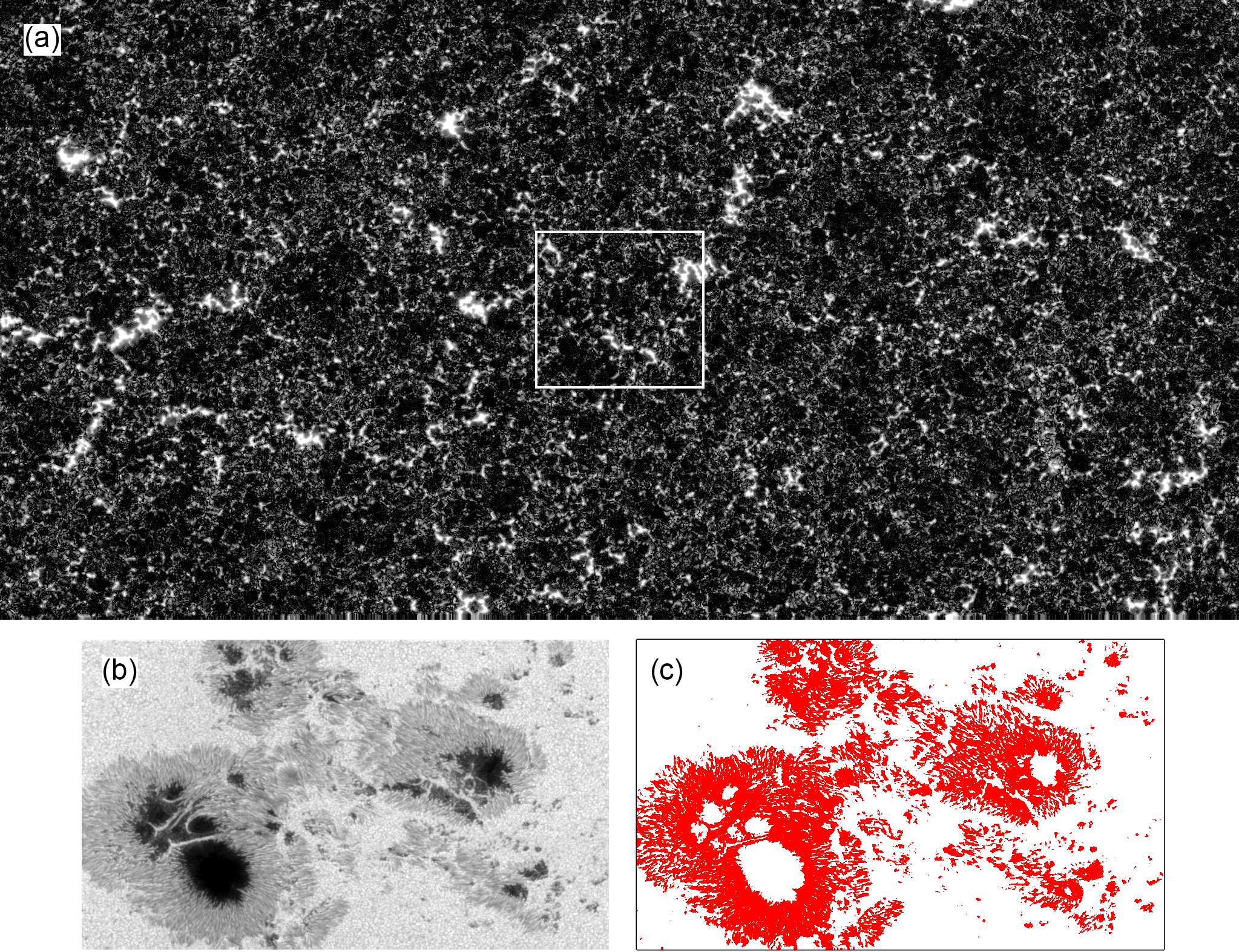}
  \caption{The maps of the areas under study. (a) -- the map of
  $\log_{10}(\BIV)$, black and white clipping at levels 1 and 2.5 (10 and
  310~G) for session 20070310\_1137. $\BIV$ denotes the longitudinal field
  flux density measured by the COG method, averaged over two lines.
  (b) and (c) -- the brightness map of $\Ic$ and the mask highlighting
  the penumbral regions ($\Ic = [0.3,\, 0.8]\,\IQS$) for session
  20141024\_0031, $\IQS$ is the average brightness of a quiet Sun.}
  \label{fig:2}
\end{figure*}

This work is based on data obtained by the spectropolarimeter (SP) [Lites
et al., 2013] of the solar optical telescope (SOT) [Tsuneta et al., 2008]
of the Hinode satellite mission [Kosugi et al., 2007]. We use level 1 data
that has already been preprocessed. The main data analyzed is a quiet
photosphere without dividing it into Network (NW) regions and regions
inside supergranules (InterNetwork, IN). In the present study, the quiet
photosphere is represented by an observation session, which we identify as
20070310\_1137 (see Fig.~2(a)). The first eight digits indicate the date in
YYYYMMDD format, and the last four digits indicate the hours and minutes
when the session started. This session is one of the most studied,
starting with [Lites et al., 2008] paper and ending with [Amari et al.,
2024]. As an illustration, the central part of the $40\times40$ arcsecond
map of this session will be shown in Fig.~7 below (just as it was done in 
[Lites et al., 2008]). For penumbra analysis, it is convenient to use 
session 20141024\_0031 (Figs.~2(b) and~2(c)).

The SP/Hinode data consist of the Stokes intensities $\StI$, $\StQ$, $\StU$, $\StV$
recorded for 112 points in the spectrum in the wavelength range
6300.9--6303.0~\AA{} in steps of about 21.6~\mA{}. The polarimeter can
register up to 1024 points along the slit in $0.16''$ increments, and the
slit can also move along the surface in up to 2047 steps in $0.15''$
steps. The SOT telescope limits its angular resolution to $0.32''$, which
is related to its diffraction limit. This paper uses maps recorded with an
exposure of 4.8~s for each slit position.

When studying the magnetic field of the quiet photosphere, one must first
discard map points that do not show the magnetic splitting of the lines.
We assume that these are the points for which the values of the Stokes
parameters $\StQ$, $\StU$, and $\StV$ are in total less than a certain threshold
value. We define the polarization value as $P=\sqrt{\StQ^2+\StU^2+\StV^2}$, where
the values of $\StQ$, $\StU$ and $\StV$ are normalized by the continuum level
$\Ic$. By summing up the $P$ values along the line profiles, we can
determine the equivalent width of the polarization $\WP$ by analogy with
the equivalent width of the spectral line derived from the intensity
profile $\WI = \int_{\lambda_1}^{\lambda_2}(\StI/\Ic)\,d\lambda$. The value of
$\WP$ can be described as the area between the polarization curve and the
zero axis. Since the $Y$-axis is dimensionless, $\WP$ has the wavelength
dimension, as well as the $X$-axis. We determine the values of the
summation boundaries $\lambda_1$ and $\lambda_2$ by one of two criteria
-- either the intensity in the wing of the profile $\StI(\lambda)$ becomes
less than 95\%~$\Ic$, or the distance from the center of the line
$\lambda_0$ exceeds 300~\mA{}. A histogram was constructed for the values
of $\WP$ from all $2\times10^{6}$ points on the map. If we assume that
most points in the quiet photosphere do not contain a magnetic field, then
the position of the top of the histogram will indicate the average value
of the noise level. It is 1.6~\mA{}. When plotting the scatter plots, we
will limit ourselves from below to the value $\WP = 3.3$~\mA{}, thus
discarding 92.5\% of the non-magnetic map points for the session
20070310\_1137 (as in Figs.~1(a,\,b,\,c) for example).

Values $\WP$ are important for estimating polarization in lines. However,
to determine the noise level, the $\Ptot$ value (``the fractional total
net polarization'') used by [Lites et al., 2008] is more convenient. Our 
approach to determining $\WP$ is identical to that of these authors, 
the difference lies in a constant coefficient, $\Ptot = \WP/1200$. The
level that separates magnetic and non-magnetic points $\Ptot = 0.004$
adopted in their work corresponds to $\WP = 4.8$~\mA{}. The $\Ptot$ and
$\WP$ relation is explained by the fact that the authors [Lites et al.,
2008] divided the integral by the length of the integration sections. The
values of $\lambda_1$ and $\lambda_2$ were chosen by these authors from
$-300$ to $+300$~\mA{} for each of the two lines, for a total of
$(300+300)\times 2 = 1200$.

The values of the longitudinal field for the COG method are obtained by
summing the Stokes profiles $\StI$ and $\StV$ according to the formula~(2) from
Section~3. The flux density $f\cdot\BMEpar$ is obtained by multiplying
three physical quantities extracted from the Level~2 FITS-files: the field
strength $|B|$ (\Id{Field\_Strength}), the cosine of the magnetic
field inclination angle to the line of sight $\cos(\gamma)$ ($\gamma$ --
\Id{Field\_Inclination}), and the magnetic filling factor $f$ (which
is historically labeled as \Id{Stray\_Light\_Fill\_Factor} in the
standard Hinode SP FITS headers).

\section{Center of gravity method}

The center of gravity method is described in [Semel, 1967] and discussed
in more detail in [Rees \& Semel, 1979]. It is based on the fact that the
difference in the positions of the center of gravity (COG) $\StI+\StV$ and
$\StI-\StV$ of the Stokes profiles is equal to the value of the magnetic
splitting of the spectral line according to the formula (here the values
of $\lambda$ are given in \AA{}, and $B$ in Gauss):
\begin{equation}
  \frac{\lambda_{\StI+\StV}-\lambda_{\StI-\StV}}{2}
  = 4.6686\times10^{-13}\cdot\lambda^{2}\cdot g_{\mathrm{eff}}\cdot\Bpar ,
\end{equation}
the wavelengths of the centers of gravity, $\lIV$, where
$\IIV{} = \Ic - (\StI\pm \StV)$, are calculated as sums:
\begin{equation}
  \lIV = \frac{\sum\left(\IIV{}\cdot\lambda\cdot\Delta\lambda\right)}
              {\sum\left(\IIV{}\cdot\Delta\lambda\right)} ,
\end{equation}
$\Bpar$ from~(1) is the longitudinal component of the magnetic field
calculated by the COG method, as synonyms in the text, we use the notation
$\BIpmV$, or simply $\BIV$. At the other hand, $\Bpar = |B|\cos(\gamma)$,
where $\gamma$ is the angle between the magnetic field vector and the line
of sight.

Formulas (1) and (2) can be written in an alternative formulation. In this
case, the method is referred to as the method of moments for the Stokes
$\StV$ parameter [Mathys, 1988]. To be more precise, this is the first moment
of the Stokes $\StV$ parameter, normalized to the equivalent width of the
spectral line:
\begin{equation}
  M_{\StV} = \frac{\displaystyle\int_{\lambda_1}^{\lambda_2}
          \left(\lambda-\lambda_0\right)\cdot \StV\cdot d\lambda}{\WI} .
\end{equation}

The moment $M_{\StV}$ in the formula~(3) is physically equivalent to the
difference in the positions of the centers of gravity $\lIV$ from the
formula~(2). The results of the field calculation using formulas (2) and
(3) coincide. The form of notation~(3) allows us to understand that with a
small filling factor (Filling Factor -- $f$), meaning a low fraction of
the area in the spectrograph's aperture occupied by regions with a
magnetic field, the result may be distorted. Namely, the regions of the
surface responsible for the formation of the Stokes profiles $\StV(\lambda)$
and $\StI(\lambda)$ may not coincide. This will alter the values of $M_{\StV}$
for the \FeI{} $\lambda$6301 and 6302~\AA{} lines by different amounts.
Let us explain this.

\begin{figure*}[!t]
  \centering
  \includegraphics[width=\textwidth]{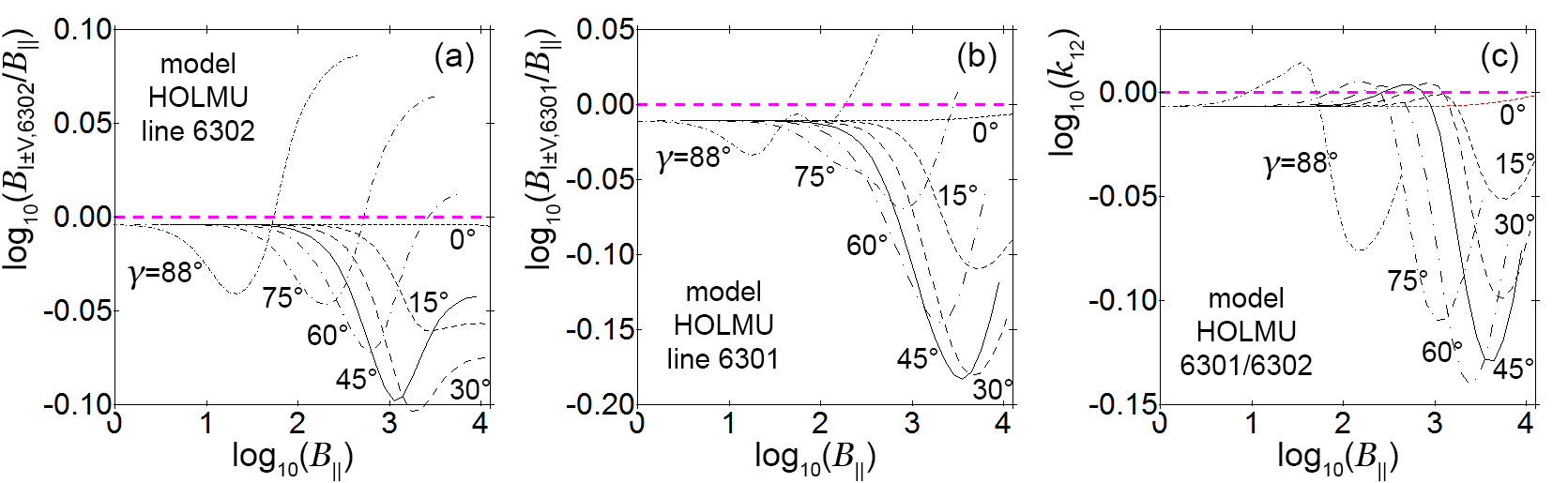}
  \caption{Comparison of the values of the longitudinal field strength.
  (a) and (b) -- comparison of the field $\BIpmV$ obtained by calculation
  from synthetic Stokes profiles with the model field
  $\Bpar = B_{0}\cdot\cos(\gamma)$ for the \FeI{} $\lambda$6302 and
  6301~\AA{} lines. (c) -- a comparison of the calculated fields for the
  two lines with each other. The OX axis shows the logarithm of the field
  specified in the model in Gauss. The OY axis shows the logarithm of the
  ratio of the field values. Synthetic Stokes profiles were calculated
  using the HOLMU photosphere model [Holweger \& Müller, 1974] for 7
  values of the angle of inclination of the field vector to the line of
  sight and a set of field strength values. $\log_{10}(0.05, 0.1, 0.15,
  0.2)$ corresponds to 1.12, 1.26, 1.41, 1.58.}
  \label{fig:3}
\end{figure*}

So, the presence of thin magnetic flux tubes with kG field strength can
change the relation of the longitudinal field values
$\kt=\BIVl{6301}/\BIVl{6302}$ described in the Introduction.
Schematically, the distortion mechanism can be described as follows. It
is known that the equivalent widths of the lines weaken in regions of the
quiet photosphere with a magnetic field outside of sunspots. Let the field
$B_{0}=1000$~G, the fill factor $f=0.1$ and the equivalent width for
\FeI{} $\lambda$6302~\AA{} line is reduced by half $r_{2}=0.5$. Here, $r$
is the coefficient of change of the equivalent width of the line in the
magnetic field relative to the non-magnetic photosphere. Then the measured
flux density is $\BIVl{6302}=B_{0}\cdot f\cdot r_{2}=50$~\flux{}. If
$W_{\StI,6301}$ decreases in the magnetic field by a factor of 1.5,
$r_{1}\approx0.7$, then $\BIVl{6301}$ will be equal to 70~\flux{}. As a 
result, the ratio of the measured longitudinal field strengths $\kt=1.4$. 
It is clear that the values of $B_{0}$ and $f$ for both lines are the same;
hence, $\kt$ can be simply calculated as $r_{1}/r_{2}$ (when $f\ll 1$). In
the paragraph above, the specific values of $r_{1}=0.7$ and $r_{2}=0.5$
were taken merely for illustration. To estimate $r_{1}$ and $r_{2}$, we
can use the $\WI$ values obtained from observations for the session map
20070310\_1137 for two statistical subsets of points. One subset consists
of map points without a field, filtered by $\WP < 3$~\mA{}. The second
subset consists of points with $\BME$ (\Id{Field\_Strength}) values greater
than 900~G and a filling factor $f$ greater than 0.75. The obtained values 
are presented in the table:

\begin{strip}
  \centering
  \small
  \setlength{\tabcolsep}{6pt}
  \renewcommand{\arraystretch}{1.25}
  \begin{tabular}{|l|c|c|c|}
    \hline
    The Line / Physical Quantity & $\WI$ ($B=0$), \mA{}
      & $\WI$ ($\BME>900$~G, $f>0.75$), \mA{}
      & $r=W_{\StI,B>900}/W_{\StI,B=0}$ \\
    \hline
    6301 & 126.5 & 97.4 & 0.694 \\
    \hline
    6302 & 87.8 & 65.0 & 0.667 \\
    \hline
  \end{tabular}
\end{strip}

Comparing $r_{1}$ and $r_{2}$, we find that $\kt\sim 1.05$. This implies
that either the kG thin tubes cannot provide the observed average value of
$\kt = 1.3$, or $r_{1}/r_{2}\gg 1.05$, so the model of the photosphere for
these tubes must differ significantly from the model of the photosphere
with homogeneous magnetic fields ($f = 1$) at the same field strength.

The method of centers of gravity, which is equivalent to the method of
Stokes $\StV$ parameter normalized moment, has important beneficial
properties. The difference in the positions of the centers of gravity
$\lIV$ is linearly related to the strength of the longitudinal component
of the magnetic field. This linearity has important consequences:
\begin{enumerate}
  \item The calculation result will not change when the photosphere model
        is changed.
  \item It should work equally well for any Zeeman-sensitive line.
\end{enumerate}

In other words, by taking measurements in the umbra, penumbra, and areas
of quiet photosphere, we should obtain uniform and accurate results. We
will also achieve consistent results when measuring in different spectral
lines, particularly in the \FeI{} $\lambda$6302 and 6301~\AA{} lines.

It is also important to note that in the process of integration by the
formula~(3), the noise-induced distortions of the $\StV$ values are averaged.

\begin{figure*}[!t]
  \centering
  \includegraphics[width=109mm]{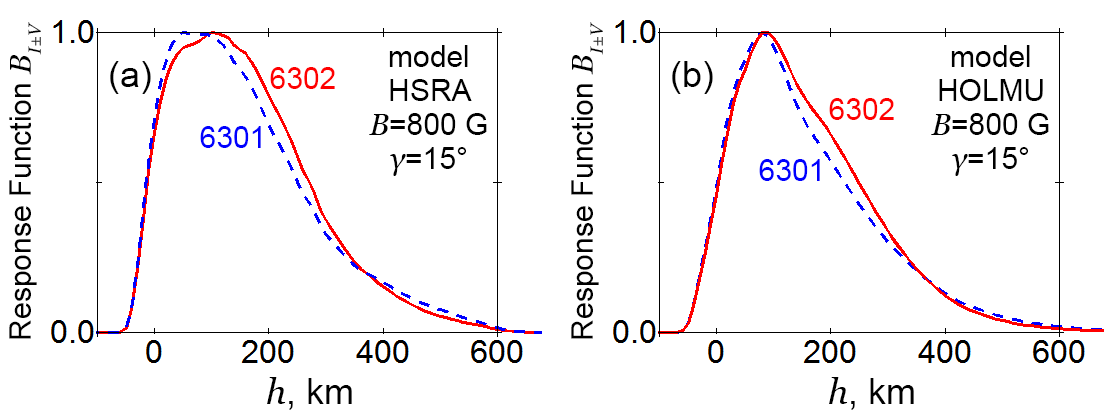}
  \caption{Calculation of response functions RF for the lines \FeI{}
  $\lambda$6301 and 6302~\AA{} using one-dimensional models of the
  photosphere HSRA [Gingerich et al., 1971] and HOLMU [Holweger \& Müller,
  1974]. In a layer with a thickness of $\Delta h = 10$~km, a disturbance
  of the magnetic field strength $B$ with a magnitude of 10~G was applied
  successively at each heights of the photosphere. Stokes profiles were
  constructed, and the response of $\BIpmV$ was calculated from them.}
  \label{fig:4}
\end{figure*}

In the course of current research, a series of Stokes profile calculations
were performed for a one-dimensional photosphere model, specifically using
the HOLMU model [Holweger \& Müller, 1974]. The calculation program
[Mozharovsky, 2013] is a functional analogue of the MALIP [Landi
Degl'Innocenti, 1976] and SPANSAT [Gadun \& Sheminova, 1988] codes. The
calculations revealed the following important properties of the COG
method:
\begin{enumerate}
  \item The results of calculations are independent of the convolution of
        the Stokes profiles with the instrumental function of the
        spectrograph or with Doppler broadening due to macro- and
        microturbulence. Moreover, even extremely large linear gradients
        of line-of-sight (LOS) velocity do not significantly affect the
        result as long as the field strength does not change with height.
  \item Also, the result does not depend much on the step of the profiles
        along the wavelength, that is, for measurements with a step of,
        for example, 10, 20 and 30~\mA{}, the calculation results will be
        approximately the same.
\end{enumerate}

In calculations for weak lines (equivalent widths $\WI < 15$~\mA{} at
wavelength $\sim 6000$~\AA{}) formula~(1) remains valid with an accuracy of
a few percent. The values of $\BIV$ are underestimated by tens of percent
for strong lines, that is, lines with larger equivalent widths.

The properties of the COG calculation results for the \FeI{} $\lambda$6302
and 6301~\AA{} lines obtained by numerical calculations for the
one-dimensional photosphere model are shown in the graphs (Fig.~3). As
these calculations have shown, the COG method usually underestimates the
values of the longitudinal field. The degree of underestimation depends on 
the angle of the field with respect to the line of sight $\gamma$, on the 
field strength, and on the spectral line itself. In particular, for
$\gamma=0\degr$, the COG method works without distortion, but it is obvious
that real fields are rarely directed exactly in the direction of the
observer. In our 1D model calculations, for the \FeI{} $\lambda$6301~\AA{}
line, the values of $\BIV$ are systematically less than for \FeI{}
$\lambda$6302~\AA{}, $\kt=B_{6301}/B_{6302}<1$. For normal fields of a quiet
Sun, $\kt\sim 0.98$, but $\kt$ values can also reach 0.8--0.9 at field
strengths greater than $10^{3}$ Gauss. This cannot in any way explain the
observed average strength difference in the quiet photosphere $\kt = 1.3$
(see Fig.~1(c)). On the other hand the differences between the field values
in the penumbra obtained using ME inversion and the COG method
(Fig.~1(e)) $\kme=1.1$--1.2 can be explained solely by the underestimation
of values using the COG method.

It is important to note that the data obtained using the COG method for
the umbra of the spot -- specifically for regions with a continuous
spectrum level $\Ic < 0.3\,\IQS$ (QS -- Quiet Sun) -- are unreliable due
to the large number of blending lines, both molecular and atomic. These
lines themselves exhibit polarization in the magnetic field and,
furthermore, do not allow for the determination of the continuous spectrum
level, thus making it impossible to calculate the equivalent width $\WI$,
which is included in the formulas for calculating $\BIV$.

\section{The ratio of \texorpdfstring{$\BIVl{6301}/\BIVl{6302}$}{BIV,6301/BIV,6302} strengths and the
difference in response heights for two lines}

As stated in the Introduction, according to Fig.~1(c), the longitudinal
field measured in the \FeI{} $\lambda$6301~\AA{} line in the quiet
photosphere exceeds the field measured in the \FeI{} $\lambda$6302~\AA{}
line by an average of 30\% ($\kt = 1.3$) for $\BIV$ values of
50--80~G and decreases to $\kt = 1.15$ at field strengths of
350~G. One possible explanation is that the longitudinal field
varies with height, and the response functions for COG measurements for
different lines have different effective heights. The commonly observed
increase in the field with height in the magnetic elements of a quiet
photosphere contradicts the generally accepted concept of magnetic flux
tube expansion. Therefore, we need to find conditions under which the
response height for the \FeI{} $\lambda$6301~\AA{} line is deeper than for
\FeI{} $\lambda$6302~\AA{}.

\begin{figure*}[!t]
  \centering
  \includegraphics[width=\textwidth]{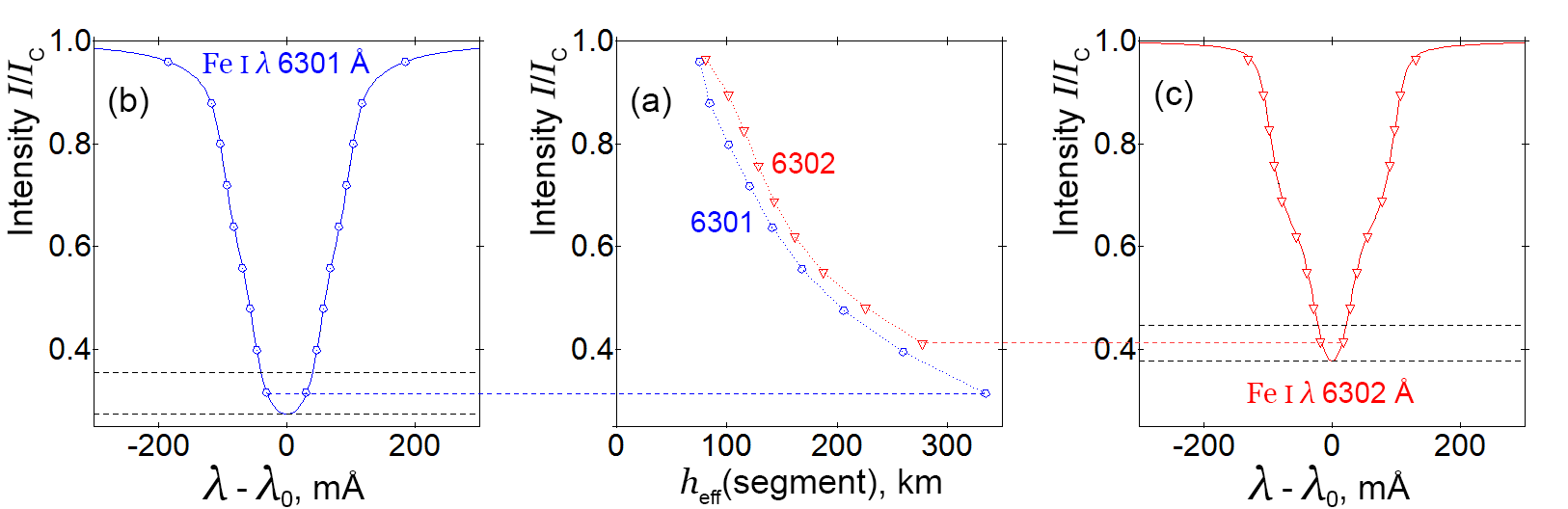}
  \caption{Comparison of effective heights of response for different
  segments of the spectral line intensity profile. For each point on the
  graph (a), its position on the scale $Y=\StI/\Ic$ indicates the
  corresponding segment of the profile $\StI$ on (b) and (c).}
  \label{fig:5}
\end{figure*}

In order to analyze the response heights, we performed a series of
calculations of the effective response heights of $\BIpmV$ to changes in
the magnetic field strength $B$. The calculations were performed
numerically. In this numerical method, a series of calculations is
performed with different heights of a certain ``test'' (or ``probe'')
layer, where a certain physical parameter in the height interval $\Delta h$
is given a small increment, for example, $B = B_{0}+10$~G. For the value
$\BIpmV$ calculated from the formula~(2), we find the variations, that is,
the Response Function -- RF ($\BIpmV$), and plot it on the graph
RF($h$), where $h$ -- height, see Fig.~4. We take the position of the 
center of gravity of the response function on the $h$ scale as the effective
response height. Calculations have shown that the shape of RF functions
practically does not depend on the field strength (at photospheric values,
i.e. $B \le 1000$~G), and also shows little dependence on the angle of the
field vector to the line of sight $\gamma$. There is a dependence on the
photosphere model, see Figs.~4(a) and~4(b). However, it does not change the
height relationship for the \FeI{} $\lambda$6301 and 6302~\AA{} lines. As
it turned out, the response $\BIpmV$ for the \FeI{} $\lambda$6301~\AA{}
line is formed deeper than the response for \FeI{} $\lambda$6302~\AA{},
although the height difference is very small. It is commonly known that
stronger lines are formed at greater heights than weaker ones. But it
turns out that this may not be true if we define a response height that
requires integrating the entire profile, not just the central part of it.
We divided the profiles $\StI\pm \StV$ into 9 equal intervals along the residual
intensity $\StI/\Ic$ of each of the lines and repeated the calculation of the
effective heights, see Fig.~5. The central part of the profile of the
\FeI{} $\lambda$6301~\AA{} line, as expected, is formed at the highest
levels. But each individual profile fragment with the same residual
intensity is formed for line 6301 deeper than for line 6302. However, the
difference in line response heights is small. To explain the difference in
the measured field strength $\kt = 1.3$, it is necessary to assume
sufficiently large gradients of the longitudinal field. How can the
longitudinal component of the field change rapidly with height? Probably
due to its bending from a vertical position to a horizontal position. One
may assume that the abundant horizontal fields $B_{T} = 55$~\flux{}
[Lites et al., 2008] are carried away by convective motion into
intergranular lanes, where they are drawn into deeper layers and quickly
turn into vertical (longitudinal).

Let us introduce the following field configuration into the HOLMU
photospheric model [Holweger \& Müller, 1974] -- the absolute value of
the field in the model $B_{0}$ remains constant. The angle of inclination
of the field to the line of sight in the deeper layers is $0\degr$, while
in the upper layers it is $90\degr$, and over a certain interval $\Delta
h_{t}$ (where $h_{t}$ is the Height of Turn), it changes linearly with
height, with the center of this interval occurring at the height $h_{t}$.
We calculate the profiles and compare the values $\BIVl{6301}$ and
$\BIVl{6302}$. In this simple way, it was possible to obtain values for
$\kt$ up to 1.2, see Fig.~6(a). However, such values are obtained under
certain conditions. On both sides of a certain peak near $B_{0}=900$~G,
the value of $\kt$ significantly decreases. The reduction in the height
range of the turn $\Delta h_{t}$ increases $\kt$. However, even with
$\Delta h_{t} = 40$~km, the best value of $\kt$ is still close to 1.2. The
model also allows us to calculate the limit (physically unreal) case
$\Delta h_{t} = 0$, which allows us to consider the limit values
(Figs.~6(b) and~6(c)). If we change the average height of the turn (within
$h_{t} = 0$--100~km above the level of $\tau_{5} = 1$), then this does not
significantly change $\kt$, but it greatly changes the apparent
longitudinal field strength. To get the most frequent statistical observed
values $\BIV\sim50$--80~G the turn height $h_{t}$ for the HOLMU model
should fall in the range from 20 to 80~km. If it is higher, we will obtain
observed longitudinal fields $\BIV\sim200$--300~G; if lower, about 10~G.
This suggests that in the Network region, the optical depth $\tau_{5}$, at
which we observe the ``turn,'' may change. This leads to both the
measurable strengths $\BIV\sim300$--800~G and the fact that the values of
$\kt$ drop to~1.

\begin{figure*}[!t]
  \centering
  \includegraphics[width=\textwidth]{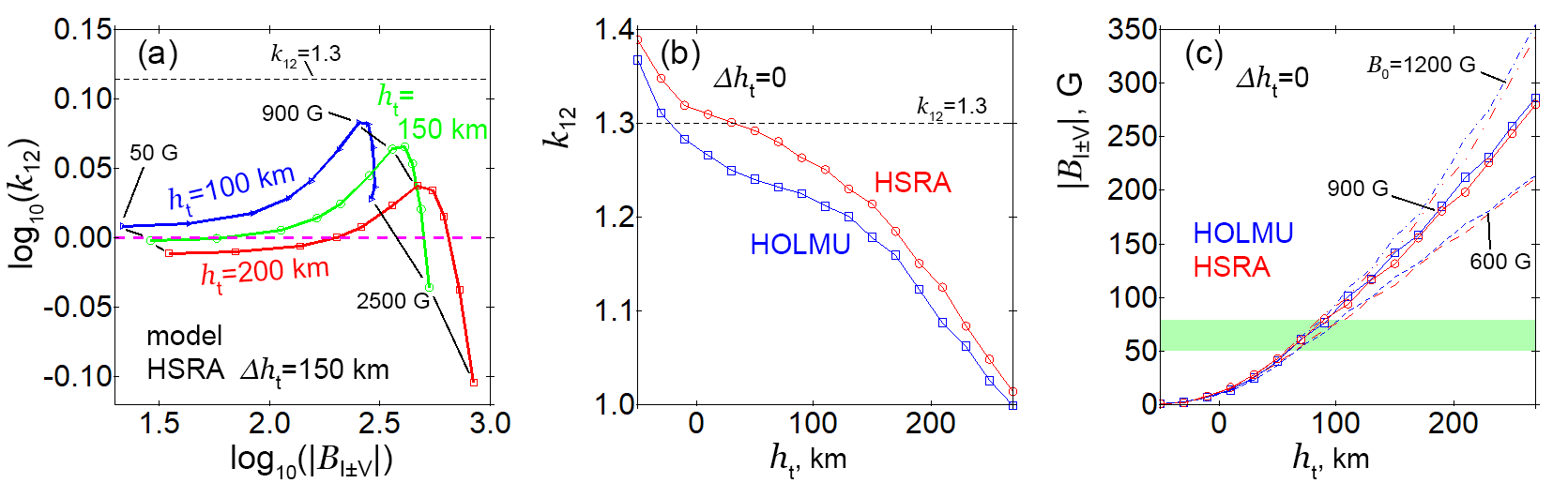}
  \caption{Model calculations of the magnetic field turn from the vertical
  direction to the horizontal one. In the graphs, $h_{t}$ is the average
  Height of the Turn, $\Delta h_{t}$ is the distance from the start to the
  finish of the turn, $|\BIpmV|$ is the longitudinal field strength
  obtained from the $\StI$ and $\StV$ Stokes profiles calculated from the model,
  $B_{0} =$ the strength, specified in the model, $\kt$ is the ratio of the
  strengths $\BIpmV$ calculated for the \FeI{} $\lambda$6301 and
  6302~\AA{} lines. Figure (a) -- the variation of $\kt$ for a series of
  $B_{0}$ values from 50 to 2500~G; the field was rotated $90\degr$ at an
  altitude interval of 150~km in three positions of the center of the turn
  in height $h_{t}$. Figures (b) and (c) -- calculations in which the
  field turned at a single point ($\Delta h_{t} = 0$), a series of values
  $h_{t}$ is given, and the dependence on the photosphere model is viewed.
  In particular, (b) -- changes in the value of $\kt$ at $B_{0} = 900$~G
  and (c) -- changes in the apparent longitudinal field $\BIV$ at
  $B_{0} = 600$, 900 and 1200~G.}
  \label{fig:6}
\end{figure*}

\begin{figure*}[!t]
  \centering
  \includegraphics[width=0.6667\textwidth]{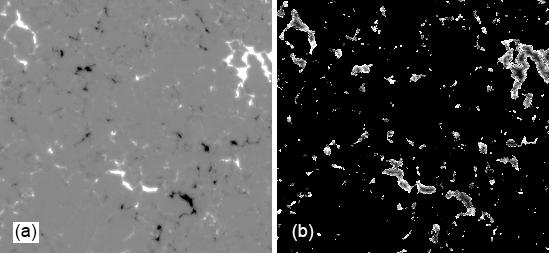}
  \caption{The central fragment of size $40\times40''$ for session
  20070310\_1137 map. (a) -- $\BIV$ magnetogram, cut-off levels $\pm
  300$~G. (b) -- map of $\log_{10}(\kt)$, cut-off levels $[0,\, 0.2]$
  ($\kt\sim[1,\, 1.58]$); the black mask covers areas with a magnetic
  field close to zero, for which $\WP < 3.3$~\mA{}.}
  \label{fig:7}
\end{figure*}

It should be noted that all the calculations made above assume a
one-component one-dimensional model of the photosphere. That is, the
change in the photosphere in the plane is significantly less than its
changes in the vertical direction. The second remark is that the
calculations do not include LOS velocities, which, in the presence of
vertical gradients, can significantly alter the effective response heights
in the wings of the line profiles. Therefore, the obtained values can be
considered as a preliminary conditional estimate. What can be said more
definitively is that the magnetic structures in the quiet photosphere are
morphologically homogeneous. Any magnetic field hill at the periphery is
surrounded by a region in which the longitudinal field is weaker than in
the center. As we can see on the map in Fig.~7(b), these peripheral regions
have significant values $\kt$ (on average, about 1.3), relatively small
measured fields $\BIV\sim50$--80~G, and a significant weight in statistics.
As one of the explanations for the large values of $\kt$, we can assume
that here we observe a change of inclination of the magnetic field lines
from the horizontal to the vertical direction as the field lines are drawn
into the lower layers of the photosphere by descending convective flows.

\section{Conclusions}

The method of centers of gravity (COG) allows us to expect almost
identical values of the longitudinal field measured using the \FeI{}
$\lambda$6301 and 6302~\AA{} lines if the field is uniform in height and
within the aperture of the spectrograph. In practice, the ratio of
strengths in these lines everywhere in the quiet photosphere is noticeably
greater than unity, on average $\kt = 1.3$. For researchers, this remains
a puzzle, as noted in the review [Bellot Rubio \& Orozco Suárez, 2019]. It
is clear that the explanation lies in the fact that these two lines
reflect the state of the field strength in different volume elements of
the photosphere. Traditionally, since [Stenflo, 1973], it has been assumed
that the quiet photosphere is permeated by ubiquitous thin flux tubes of
very small diameter. The presence of these tubes somehow affects the
results of all magnetic field measurements. In particular, some authors
explain the observed strength ratio $\kt = 1.3$ within this approach
[Domínguez Cerdeña et al., 2003; Botygina et al., 2017].

However, in the framework of this work, another possibility is proposed.
Namely, the effect on the measurement results of field inhomogeneities not
in the plane of the aperture, but along the line of sight. The current
study demonstrates that the commonly accepted notion that a stronger line
always reflects the physical properties of the photosphere from greater
heights is incorrect. In particular, the response to a change in the
$\BIV$ field strength for the \FeI{} $\lambda$6301~\AA{} line originates
deeper than the response of the \FeI{} $\lambda$6302~\AA{} line. However,
the difference in response heights is small, which requires a sufficiently
sharp change in the magnitude of the longitudinal field with height to
explain the observed ratio $\kt = 1.3$. As a result of model calculations,
it was shown that the observed ratio can be partially explained by
assuming a sharp change in the slope of the magnetic field lines with
height from an almost vertical field at greater depths to an almost
horizontal one in the upper layers. The interval at which changes occur
should be quite narrow -- tens of kilometers. The height in the
photosphere at which such a change occurs also has a certain value, it
must be in the range from 0 to 100~km above the level $\tau_{5} = 1$, so
that the measured longitudinal field $\BIV$ is in the range of the most
statistically probable values 50--80~\flux{}. This picture is somewhat
consistent with the observed excess of horizontal magnetic flux of
55~\flux{} over vertical flux of 11~\flux{} found in the InterNetwork
areas [Lites et al., 2008]. Also, two studies [Stenflo, 2013; Lites et al.,
2017] investigated the orientation of the polarization plane near the
limb. From these studies, it follows that at certain height, the field is
mostly horizontal relative to the Sun's surface, but when moving to deeper
layers, the field becomes more vertical. Finally, the distribution of
$\kt$ on the Hinode session map for the quiet photosphere (Fig.~7(b))
shows its increase along the periphery of the magnetic field hills, which
may indicate a sharp change in the vertical direction of the field in
these places.

\subsection*{FINANCING OF THE WORK}
This work was funded from the Institute's budget. No additional grants
were received to conduct or direct this particular study.

\subsection*{CONFLICT OF INTEREST}
The author declares that he has no conflict of interest.

\section*{REFERENCES}
\addcontentsline{toc}{section}{References}

\begin{journalrefs}

\item Amari T., Canou A., Velli M. et al. The Ubiquity of Twisted Flux
Ropes in the Quiet Sun // arXiv:2411.10563. 2024.
DOI: 10.48550/arXiv.2411.10563.

\item Bellot Rubio L., Orozco Suárez D. Quiet Sun magnetic fields: an
observational view // Living Reviews in Solar Physics. V.~16. P.~1. 2019.

\item Botygina O., Gordovskyy M., Lozitsky V. Investigation of Spatially
Unresolved Magnetic Field Outside Sunspots Using Hinode/SOT Observations
// Astroinformatics. M.~Brescia. V.~325. P.~59--62. 2017.

\item Domínguez Cerdeña I., Sánchez Almeida J., Kneer F. Inter-network
magnetic fields observed with sub-arcsec resolution // Astronomy and
Astrophysics. V.~407. P.~741--757. 2003.

\item Gadun A.S., Sheminova V.A. SPANSAT: the Program for LTE Calculations
of Absorption Line Profiles in Stellar Atmospheres // Preprint ITF-88-87R.
Institute of Theoretical Physics of the Academy of Sciences of the
Ukrainian SSR: Kiev. P.~37. 1988. (In Russian).

\item Gingerich O., Noyes R.W., Kalkofen W., Cuny Y. The
Harvard-Smithsonian reference atmosphere // Solar Physics. V.~18(3).
P.~347--365. 1971.

\item Holweger H., Müller E.A. The photospheric barium spectrum: solar
abundance and collision broadening of Ba~II lines by hydrogen // Solar
Physics. V.~39. P.~19--30. 1974.

\item Kosugi T., Matsuzaki K., Sakao T. et al. The Hinode (Solar-B)
Mission: An Overview // Solar Physics. V.~243(1). P.~3--17. 2007.

\item Landi Degl'Innocenti E. MALIP --- a programme to calculate the Stokes
parameter of magnetoactive Fraunhofer lines // Astronomy and Astrophysics
Supplement Series. V.~25. P.~379--390. 1976.

\item Lites B.W., Akin D.L., Card G. et al. The Hinode Spectro-Polarimeter
// Solar Physics. V.~283. P.~579--599. 2013.

\item Lites B.W., Kubo M., Socas-Navarro H. et al. The Horizontal Magnetic
Flux of the Quiet-Sun Internetwork as Observed with the Hinode
Spectro-Polarimeter // The Astrophysical Journal. V.~672. P.~1237--1253.
2008.

\item Lites B.W. et al. Are Internetwork Magnetic Fields in the Solar
Photosphere Horizontal or Vertical? // The Astrophysical Journal.
V.~835:14. 2017.

\item Mathys G. The diagnosis of stellar magnetic fields from spectral
line profiles recorded in circularly polarized light // Astronomy and
Astrophysics. V.~189. P.~179--193. 1988.

\item Mozharovsky S.G. Development of the SunWorld software. A review of
SunWorld features and methods from version 1990 to the current version //
\textit{(Solnechnaya aktivnost' i ee vliyanie na Zemlyu. Ezhegodnik UAFO
DVO)} Solar Activity and Its Impact on the Earth. Yearbook of the
Ussuriysk Astrophysical Observatory of the Far East Division, Russian
Academy of Sciences RAN. V.~15. P.~76--110. 2013. (In Russian).

\item Rees D.E., Semel M.D. Line formation in an unresolved magnetic
element --- a test of the centre of gravity method // Astronomy and
Astrophysics. V.~74. P.~1--5. 1979.

\item Semel M. Contribution à l'étude des champs magnétiques dans les
régions actives solaires // Annales d'Astrophysique. V.~30. P.~513. 1967.

\item Stenflo J.O. Magnetic-Field Structure of the Photospheric Network //
Solar Physics. V.~32. P.~41--63. 1973.

\item Stenflo J.O. Horizontal or vertical magnetic fields on the quiet
Sun. Angular distributions and their height variations // Astronomy and
Astrophysics. V.~555. P.~132. 2013.

\item Tsuneta S., Ichimoto K., Katsukawa Y. et al. The Solar Optical
Telescope for the Hinode Mission: An Overview // Solar Physics. V.~249.
P.~167--196. 2008.

\end{journalrefs}

\end{document}